\documentclass[fleqn,usenatbib]{mnras}

\usepackage{newtxtext,newtxmath}

\usepackage[T1]{fontenc}

\DeclareRobustCommand{\VAN}[3]{#2}
\let\VANthebibliography\thebibliography
\def\thebibliography{\DeclareRobustCommand{\VAN}[3]{##3}\VANthebibliography}

\usepackage{graphicx}	
\usepackage{amsmath}	
\usepackage{amssymb}	

\title[Checking high-$z$ identification of radio AGNs]{A method for checking high-redshift identification of radio AGNs}

\author[An, Zhang, Frey]{
Tao An,$^{1}$\thanks{E-mail: antao@shao.ac.cn}
Yingkang Zhang$^{1,2}$
and S\'{a}ndor Frey$^{3,4}$
\\
$^{1}$Shanghai Astronomical Observatory, Key Laboratory of Radio Astronomy, CAS, 80 Nandan Road, Shanghai 200030, China \\
$^{2}$University of Chinese Academy of Sciences, 19A Yuquanlu, Beijing 100049, China \\
$^{3}$Konkoly Observatory, Research Centre for Astronomy and Earth Sciences, Konkoly Thege Mikl\'{o}s \'{u}t 15-17, H-1121 Budapest, Hungary \\
$^{4}$Institute of Physics, ELTE E\"{o}tv\"{o}s Lor\'{a}nd University, P\'{a}zm\'{a}ny P\'{e}ter s\'et\'{a}ny 1/A, H-1117 Budapest, Hungary
}

\date{Accepted 2020 July 17. Received 2020 July 17; in original form 2020 April 7}

\pubyear{2020}

\begin{document}
\label{firstpage}
\pagerange{\pageref{firstpage}--\pageref{lastpage}}
\maketitle

\begin{abstract}
In large-scale optical spectroscopic surveys, there are many objects found to have multiple redshift measurements due to the weakness of their emission lines and the different automatic identification algorithms used. These include some suspicious high-redshift $(z \ga 5)$ active galactic nuclei (AGNs). Here we present a method for inspecting the high-redshift identification of such sources provided that they are radio-loud and have very long baseline interferometry (VLBI) imaging observations of their milli-arcsec (mas) scale jet structure available at multiple epochs. The method is based on the determination of jet component proper motions, and the fact that the combination of jet physics (the observed maximal values of the bulk Lorentz factor) and cosmology (the time dilation of observed phenomena in the early Universe) constrain the possible values of apparent proper motions. As an example, we present the case of the quasar J2346+0705 that was reported with two different redshifts, $z_{1} = 5.063$ and $z_{2} = 0.171$, in the literature. We measured the apparent proper motions ($\mu$) of three components identified in its radio jet by utilizing VLBI data taken from 2014 to 2018. We obtained $\mu_\mathrm{J1} = 0.334 \pm 0.099$~mas\,yr$^{-1}$, $\mu_\mathrm{J2} = 0.116 \pm 0.029$~mas\,yr$^{-1}$, and $\mu_\mathrm{J3} = 0.060 \pm 0.005$~mas\,yr$^{-1}$. The maximal proper motion is converted to an apparent transverse speed of $\beta_{\rm app} = 41.2\pm12.2\,c$. if the source is at redshift 5.063. This value exceeds the blazar jet speeds known to date. This and other arguments suggest that J2346+0705 is hosted by a low-redshift galaxy. Our method may be applicable for other high-redshift AGN candidates lacking unambiguous spectroscopic redshift determination or having photometric redshift estimates only, but showing prominent radio jets allowing for VLBI measurements of fast jet proper motions.
\end{abstract}

\begin{keywords}
galaxies: nuclei --- galaxies: distances and redshift --- quasars: individual: J2346+0705
\end{keywords}



\section{Introduction} 

The discovery of supermassive black holes (SMBHs) powering active galactic nuclei (AGNs) at redshifts higher than about 5, close to the end of the reionisation epoch, poses challenges for explaining the rapid growth of massive black holes in the early Universe \citep{2011MNRAS.416..216V}. Optical and near-infrared spectroscopic observations resulted in the discovery of more than 250 high-redshift ($z > 5.6$) galaxies and quasars \cite[e.g.][and references therein]{2001AJ....122.2833F,2016ApJS..227...11B,2016ApJ...833..222J,2019ApJ...873...35S}. As the observed colours of quasars depend on redshift, most high-$z$ sources were first selected as candidates using the $i$-dropout technique \citep[e.g.][]{2001AJ....122.2833F}, and then confirmed with optical spectroscopy. In the various releases of the Sloan Digital Sky Survey (SDSS) catalogue, many objects have no spectroscopic coverage, and only their photometric redshifts are given. These include a number of (candidate) high-redshift objects. Some others have suspicious or ambiguous spectroscopic redshift measurements due to the weakness of their emission lines which makes their redshift identification challenging for the automatic algorithms \citep[e.g.][]{2012AJ....144..144B,2016MNRAS.462.1603Y}. It is possible that different Data Releases (DR) of the SDSS catalogue contain two markedly different redshift values derived from the same spectrum but by different versions of the automatic pipelines.

For example, J2346+0705 (SDSS J234639.94+070506.8) is identified as a galaxy and a flat-spectrum radio source in the NASA/IPAC Extragalactic Database (NED)\footnote{\url{http://ned.ipac.caltech.edu/}}.  
Two redshifts are reported for this object in public data bases: $z_1 = 5.063$ in SDSS DR13\footnote{\url{http://skyserver.sdss.org/dr13/en/tools/explore/Summary.aspx?id=1237669517440385146}} \citep{2017ApJS..233...25A} adopted by NED, and $z_2 = 0.171$ in SDSS DR16\footnote{\url{http://skyserver.sdss.org/dr16/en/tools/explore/Summary.aspx?id=1237669517440385146}} \citep{2020ApJS..249....3A}. 
This source is possibly associated with a $\gamma$-ray source detected by the Large Area Telescope (LAT) on board the {\it Fermi} Gamma-ray Space Telescope, named as 1FHL~J2347.3+0710 \citep{2010ApJS..188..405A}, 2FHL~J2347.2+0707  \citep{2013ApJS..209...34A}, 3FGL~J2346.7+0705 \citep{2015ApJS..218...23A} and also classified as a TeV candidate \citep{2017ApJS..232...18A}.  
Until now, the most distant known $\gamma$-ray-emitting blazar is J1510+5702 at $z = 4.31$ \citep{2017ApJ...837L...5A}. As most high-$z$ blazars host SMBHs as massive as $\sim 10^9 \, \mathrm{M}_\odot$, it is important to ascertain the extremely high redshift of J2346+0705, since the value above 5 would break the high-$z$ $\gamma$-ray blazar record, possibly placing important new constraints on the growth of the first-generation SMBHs.

In radio bands, the total flux density of J2346+0705 is $\sim$200~mJy at 8.4~GHz measured with the U.S. National Radio Astronomy Observatory (NRAO) Very Large Array (VLA) \citep{2007ApJS..171...61H}, and around 230~mJy at 5~GHz measured with the NRAO Green Bank 91-m telescope \citep{1991ApJS...75....1B}. At 1.4~GHz, the NRAO VLA Sky Survey (NVSS) image shows an extended emission to the southeast on arcsec scale \citep{1998AJ....115.1693C}. 
When observed with very long baseline interferometry (VLBI) on milli-arcsec (mas) scale, the radio jet \citep{2002ApJS..141...13B,2015MNRAS.452.4274P} is characterised by a compact core and a number of knots\footnote{Astrogeo database, \url{http://astrogeo.org/cgi-bin/imdb_get_source.csh?source=J2346\%2B0705}}. The arcsec- and mas-scale jets are pointing to opposite directions.

VLBI imaging observations at multiple epochs allow us to detect positional changes and measure apparent proper motions of jet features. Based on the observed apparent proper motion--redshift ($\mu-z$) relation for a large sample of AGN jets, we introduce a method that can be applied to investigate whether certain high-redshift candidate objects are indeed at large cosmological distances, by using their radio jet proper motion measurements. In Sect.~\ref{method}, we describe the details of the procedure. We present the case of the quasar J2346+0705 as an example in Sect.~\ref{example} and discuss the possible applications and limitations of the method in Sect.~\ref{discussion}. A brief summary is given in Sect.~\ref{summary}.
In this paper, we adopt a standard flat $\Lambda$ Cold Dark Matter ($\Lambda$CDM) cosmological model with $\Omega_{\rm{m}} = 0.27$, $\Omega_{\Lambda} = 0.73$, and $H_{\rm{0}} = 70$~km\,s$^{-1}$\,Mpc$^{-1}$.

\section{The method}
\label{method}

In compact radio-emitting AGNs, components typically propagate away from the vicinity of the central SMBH along a well-defined jet. Apparent transverse speeds that reflect the bulk relativistic motion of the plasma are usually well below $\beta_\mathrm{app}=25$ (measured in the unit of the speed of light $c$) but can occasionally reach extreme values up to $\beta_\mathrm{app} \approx 40$ \citep[e.g.][]{2004ApJ...609..539K,2016AJ....152...12L,2019ApJ...874...43L}. The apparent speeds depend on the bulk Lorentz factor of the jet ($\Gamma$) and its inclination angle with respect to the line of sight ($\phi$). For a jet with a given $\Gamma$, the maximum apparent proper motion is $\beta_\mathrm{app}^\mathrm{max} = \sqrt{\Gamma^2 -1} \approx \Gamma$ \citep[see Appendix A in][]{1995PASP..107..803U}. Furthermore, the observed apparent angular proper motions measured in mas\,yr$^{-1}$ depend on the redshift of the source, because of the cosmological time dilation caused by the expansion of the Universe. It slows down phenomena in the observer's frame by a factor $(1+z)$ compared to the rest frame of the source.

\begin{figure}
\begin{center}
\includegraphics[width=0.3\textwidth,angle=270]{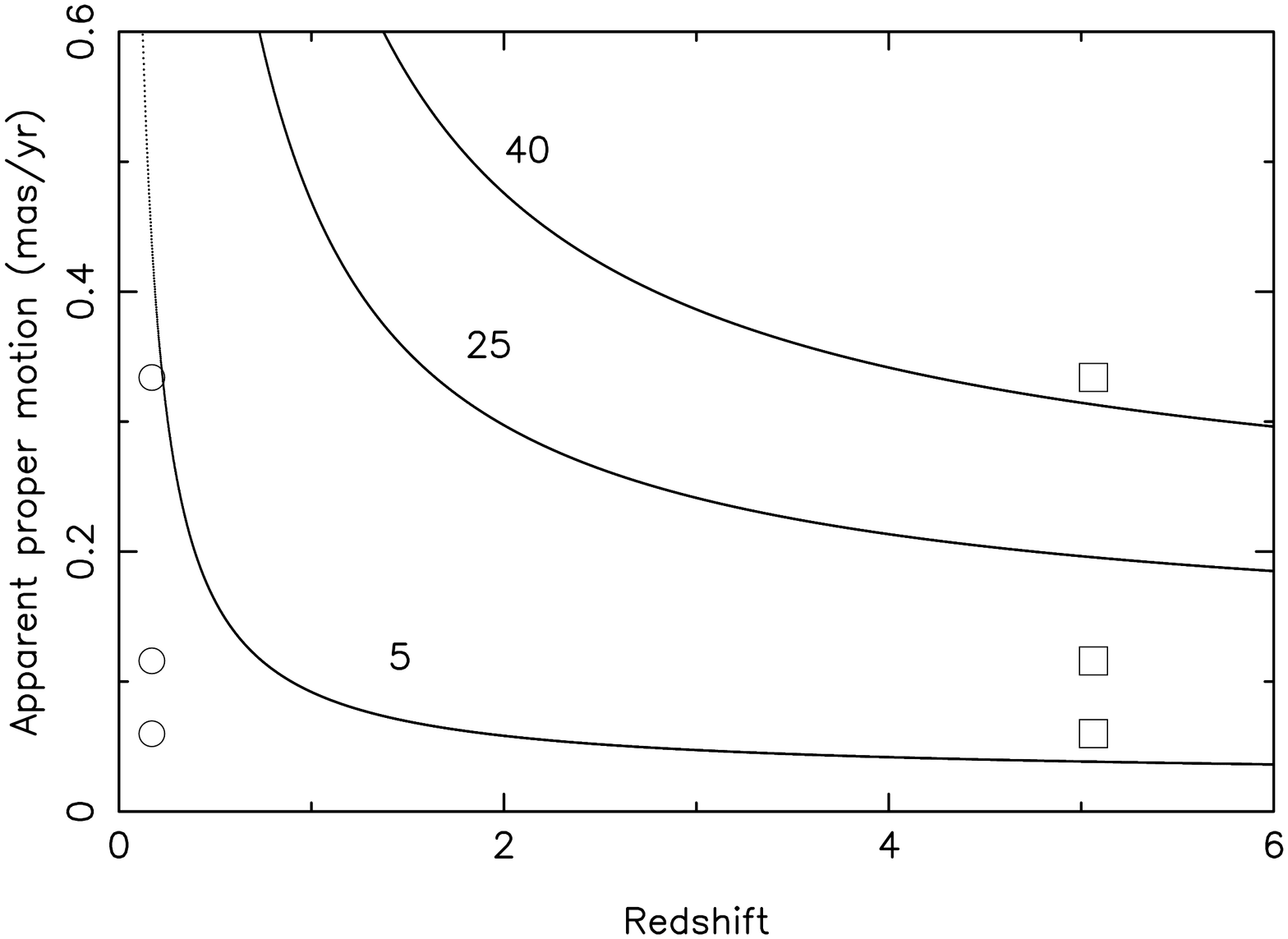}
\caption{Upper bounds of the apparent proper motion--redshift diagram for AGN jet components in the adopted $\Lambda$CDM cosmological model, assuming Lorentz factors 5, 25, and 40 (from bottom to top). The measured values for J2346+0705
are marked at $z_1=5.063$ and $z_2=0.171$ with squares and circles, respectively. At the higher redshift $z_1$, the fastest component moves more rapidly than expected from the model with the most extreme Lorentz factor, suggesting that the true redshift of J2346+0705 is  $z_2$.}
\label{fig:pm-z}
\end{center}
\end{figure}

The $\mu-z$ relation for compact radio sources was first proposed as a test of cosmological world models by \citet{1988ApJ...329....1C}. They found a clear anticorrelation between the proper motion and redshift, and a rough upper limit to $\mu$ as a function of $z$ indicating that the redshift is indeed a measure of distance. Subsequent studies of larger samples \citep[e.g.][]{1994ApJ...430..467V,2004ApJ...609..539K,2008A&A...484..119B} well established that there is indeed an upper bound in the $\mu-z$ relation that is consistent with the $\Lambda$CDM cosmology and a distribution of jet Lorentz factors with a maximum of $\Gamma \approx 25$ (see the model curves in Fig.~\ref{fig:pm-z}). 

Statistical studies of the AGN samples found that most of the apparent jet speeds are in fact lower than $5\,c$, with an extreme upper limit of $40\,c$ \citep{2016AJ....152...12L}. Consequently, an apparent transverse jet speed exceeding $40\,c$ in a distant object would naturally call its high-redshift identification in question. Alternatively, since at  $\beta_\mathrm{app} = 40$ the corresponding bulk Lorentz factor is $\Gamma \ga 40$, unprecedentedly extreme physical conditions would be required in the jet. Unless there is convincing supporting evidence for the latter, it is plausible to assume that the too fast apparent speed implies a low redshift.

\section{Application to J2346+0705}
\label{example}

Calibrated VLBI imaging data made publicly available in the Astrogeo database are used for this analysis. Data from four observing epochs are found for J2346+0705 at two frequency bands, 2.3 and 8.3/8.7~GHz. These were taken with the ten 25-m antennas of the NRAO Very Long Baseline Array (VLBA), from 1995 July to 2018 April. We choose the 8.7-GHz data that provide higher angular resolution, typically $\sim 1-3$~mas (depending on the position angle of the synthesised beam of the VLBI array) for this nearly equatorial object. The earliest data from 1995 were of relatively low quality, and thus the extended jet features cannot be reliably imaged. Therefore we restrict our analysis to the other three epochs (2014 August 6,  2017 March 23, and 2018 April 8) to estimate the apparent proper motions of the jet components identified in all images. 

\begin{figure}
\begin{center}
\includegraphics[width=0.45\textwidth,angle=0]{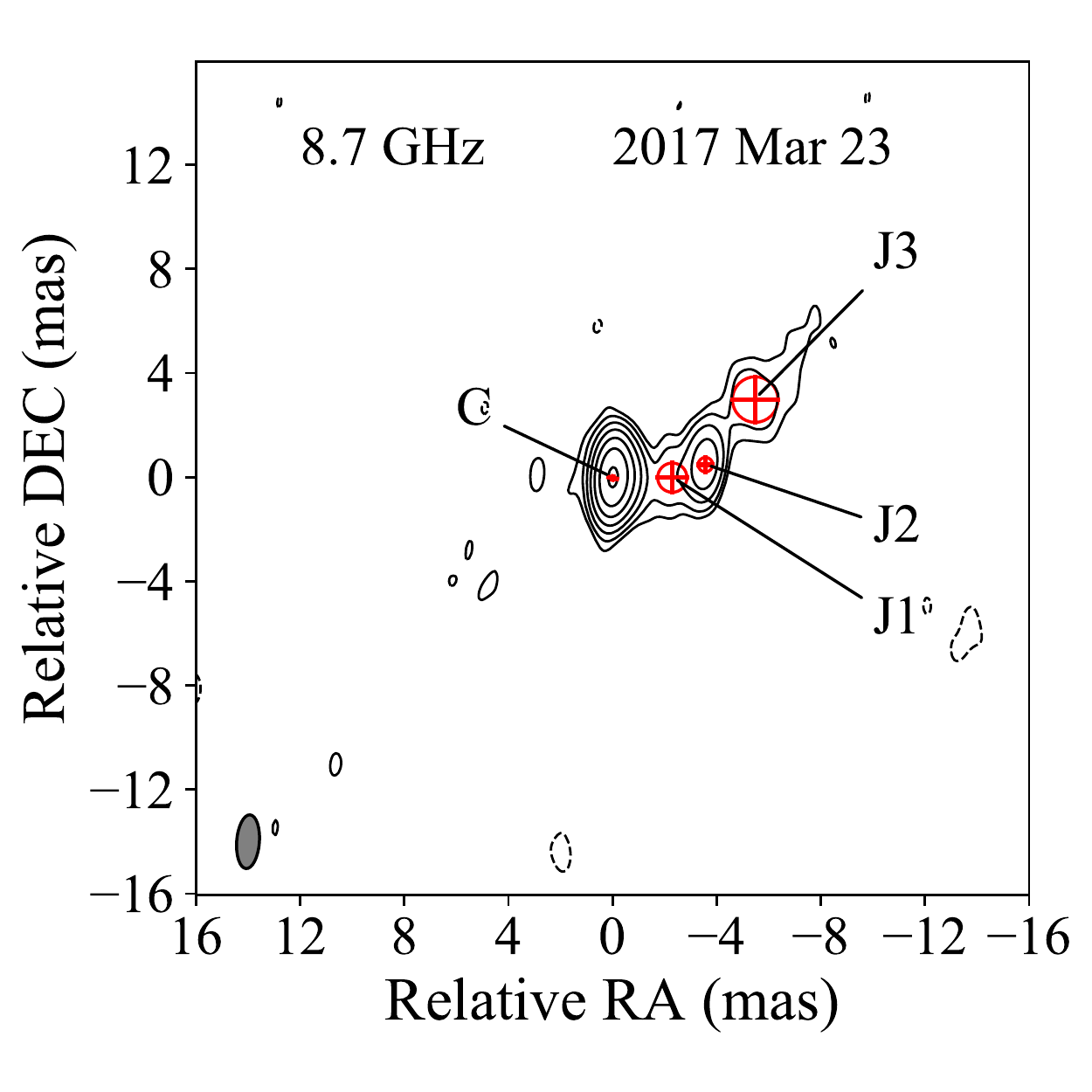}
\includegraphics[width=0.45\textwidth]{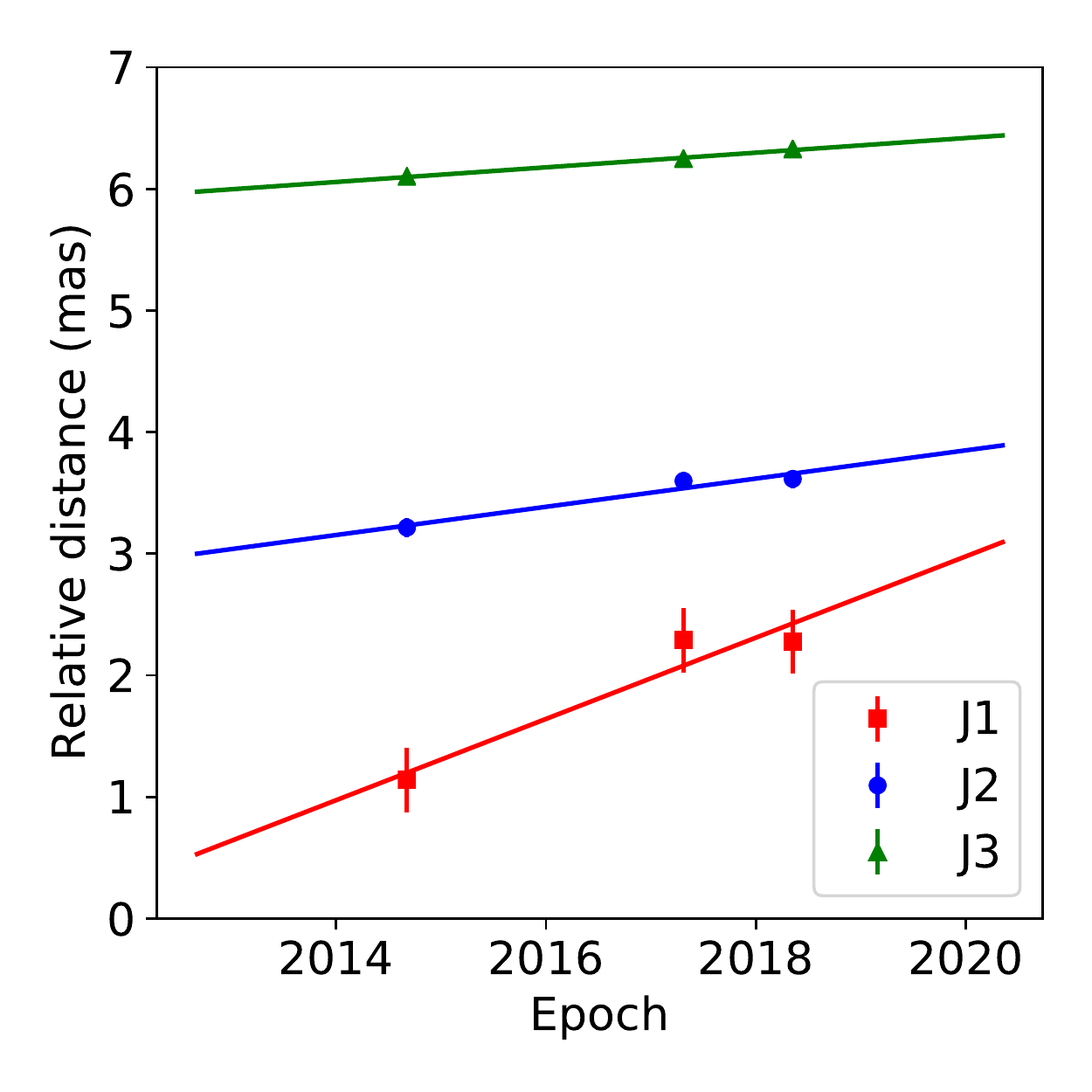}
\caption{The radio jet morphology and jet proper motion of J2346+0705. {\it Upper panel}: 8.4-GHz VLBI image on 2017 March 23. The restoring beam (shown as the grey ellipse in the bottom-left corner) is 2.1 mas $\times$ 0.9 mas (FWHM), with a major axis position angle $-4.3\degr$. The \textit{rms} noise in the image is 0.4 mJy\,beam$^{-1}$. The contours start from $\pm 1.0$~mJy\,beam$^{-1}$ and the positive levels increase by a factor of 2. The core (C) and jet components (J1, J2 and J3) are marked with red symbols. {\it Bottom panel}: separation of the jet components from the core as a function of time. The fitted linear proper motion values are $0.334\pm 0.099$ mas yr$^{-1}$ (J1), $0.116\pm 0.029$ mas yr$^{-1}$ (J2), and $0.060\pm0.005$ mas yr$^{-1}$ (J3).
\label{fig:2356}}
\end{center}
\end{figure}

Since the interferometric visibility data have already been calibrated, we only performed imaging and model fitting using the {\sc Difmap} software package \citep{1997ASPC..125...77S}. Figure~\ref{fig:2356} shows one of the total intensity images obtained from the observations. In addition to the core (C) at the image centre, there are two compact easily-recognised jet knots (J2 and J3) to the west and northwest of the core within 7~mas. An additional innermost jet component (J1) is also found in the residual map after removing the model components of the core, J2, and J3. Table~\ref{tab:modelfit} presents the parameters of the elliptical (for C) and circular Gaussian brightness distribution models fitted to the jet components in {\sc Difmap}. The model component positions and sizes are also marked in the image in Fig.~\ref{fig:2356}.

Based on the three-epoch data in Table~\ref{tab:modelfit}, we calculated the apparent proper motions of the jet components: $\mu_\mathrm{J1} = 0.334 \pm 0.099$~mas\,yr$^{-1}$ for J1, $\mu_\mathrm{J2} = 0.116 \pm 0.029$~mas\,yr$^{-1}$ for J2, and $\mu_\mathrm{J2} = 0.060 \pm 0.005$~mas\,yr$^{-1}$ for J3. The uncertainties are calculated by taking also into account the model fitting errors at the individual epochs. The apparent proper motion shows a decelerating trend, while the component sizes increase downstream the jet. This is consistent with the general picture that jet features slow down and expand when moving outwards \citep[e.g., ][]{2015ApJ...798..134H}.

\begin{table*}
    \caption{Fitted circular Gaussian model parameters for the 8.7-GHz VLBI components in J2346+0705}
    \centering 
    \begin{tabular}{cccccccc} \hline\hline
Epoch & Comp & $I$ & $S$ & $R$ & P.A. & $\theta_{\rm maj}$ & $\theta_{\rm min}$\\
(yyyy mm dd)&      & (mJy\,beam$^{-1}$) & (mJy) & (mas) & ($\degr$) & (mas) & (mas)\\
\hline
2014 08 06 & C & 96.8 $\pm$ 4.9  & 98.6 $\pm$ 5.0  & ...  & ... & 0.46 $\pm$ 0.01  & 0.14  $\pm$ 0.01   \\
           & J1 & 8.9 $\pm$ 0.6  & 10.9 $\pm$ 0.8  & 1.14 $\pm$ 0.27  & -98.1 $\pm$2.0  & 0.80 $\pm$ 0.07 & ...  \\
           & J2 & 26.2 $\pm$ 1.3  & 28.3 $\pm$ 1.5  & 3.22 $\pm$ 0.08  & -85.6 $\pm$ 0.3 & 0.89 $\pm$ 0.02 & ...  \\
           & J3 & 9.6 $\pm$ 0.6  & 14.7 $\pm$ 1.0  & 6.10 $\pm$ 0.04  & -62.3 $\pm$ 0.3 & 1.48 $\pm$ 0.07 & ... \\ 
\hline
2017 03 23 & C & 86.4 $\pm$ 4.3  & 90.4 $\pm$ 4.6  & ...  & ... & 0.32 $\pm$ 0.01 & 0.12 $\pm$ 0.01\\
           & J1 & 4.7 $\pm$ 0.5  & 8.5 $\pm$ 1.0  & 2.29 $\pm$ 0.27  & -89.7 $\pm$ 1.6 & 1.12 $\pm$ 0.12 & ... \\
           & J2 & 18.0 $\pm$ 1.0  & 19.8 $\pm$ 1.2  & 3.60 $\pm$ 0.08 & -81.9 $\pm$ 0.3 & 0.56 $\pm$ 0.03 & ... \\
           & J3 & 5.6 $\pm$ 0.5  & 12.5 $\pm$ 1.2  & 6.25 $\pm$ 0.05  & -61.2 $\pm$ 0.5 & 1.75 $\pm$ 0.10 & ...  \\ 
\hline
2018 04 08 & C & 83.9 $\pm$ 4.2  & 85.2 $\pm$ 4.3  & ... & ... & 0.34 $\pm$ 0.01 & $\le$0.07 $\pm$ 0.01 \\
           & J1 & 4.8 $\pm$ 0.4  & 9.0 $\pm$ 0.7  & 2.28 $\pm$ 0.27  & -95.4 $\pm$ 0.9 & 1.20 $\pm$ 0.07 & ...\\
           & J2 & 16.5 $\pm$ 0.9  & 20.9 $\pm$ 1.1  & 3.62 $\pm$ 0.08  & -81.1 $\pm$ 0.2 & 0.65 $\pm$ 0.02 & ...  \\
           & J3 & 6.5 $\pm$ 0.5  & 15.7 $\pm$ 1.2  & 6.33 $\pm$ 0.04  & -62.3 $\pm$ 0.4 & 1.79 $\pm$ 0.07& ...  \\ 
\hline
    \end{tabular}
\\
Notes: Col.~1 -- observing epoch, Col.~2 -- component designation, Col.~3 -- peak intensity, Col.~4 -- integrated flux density, Col.~5 -- angular separation from the core, Col.~6 -- position angle with respect t the core, measured from north through east, Col.~7 -- component diameter (FWHM).
    \label{tab:modelfit}
\end{table*}

If the redshift of J2346+0705 is $z_1 = 5.063$, then the apparent jet component speeds are $41.2 \pm12.2\,c$, $14.3 \pm 3.6 \,c$, and $7.4 \pm 0.6\,c$ for J1, J2 and J3, respectively. To date, there are only four high-redshift ($z > 4.5$) radio-loud quasars having jet proper motion measurements based on repeated VLBI imaging 
\citep{2015MNRAS.446.2921F,2018MNRAS.477.1065P,2019arXiv191212597Z,2020NatCo..11..143A}, and the values are $\lesssim 10\,c$. The apparent jet speed in the case of the innermost component J1 in J2346+0705 would be much higher than those, and in fact would approach the maximum value of $\sim50\,c$ measured for any AGN at 15~GHz \citep{2016AJ....152...12L}. Moreover, apparent jet proper motions are known to be smaller at lower frequencies \citep[e.g.][]{2008A&A...484..119B}. Note that for other $z > 4.5$ radio quasars studied so far, the jet components tend to accelerate as their distance from the core increases, possibly suggesting a young and growing jetted AGN in the early Universe \citep{2020NatCo..11..143A}.

In the light of the above observational results, we are inclined to consider J2346+0705 as a low-redshift object with $z_2=0.171$, as found in SDSS DR16 \citep{2020ApJS..249....3A}. It would solve the conflict between the anomalously fast apparent jet component motion and the high redshift in a straightforward way. Further supporting pieces of evidence against the high-redshift scenario are the $\gamma$-ray emitting nature of J2346+0705, its optical magnitudes that are brighter by at least $3^\mathrm{m}$ than those of typical $z \sim 5$ quasars \citep[see the catalogue\footnote{\url{http://astro.elte.hu/~perger/catalog.html}} of $z>4$ AGNs compiled by][]{2019MNRAS.490.2542P}, and the $g-r=0.71$ optical colour from SDSS DR16 data which is incompatible with the high redshift \citep[cf.][]{2013MNRAS.435.3306A}.

\section{Discussion}
\label{discussion}

\subsection{Jet parameters of J2346+0705}

Accepting the redshift $z_2=0.171$, we calculate the core brightness temperature, $T_\textrm{b} = 4.7 \times 10^{10}$ K, from the fitted Gaussian component sizes and flux densities (Table~\ref{tab:modelfit}). Assuming equipartition condition in the core between the particle and magnetic field energy densities \citep{1994ApJ...426...51R}, the Doppler boosting factor can be inferred as close to unity, $\delta = 0.9 \pm 0.3$. 
Adopting the apparent jet speed ($\beta_\mathrm{app} = 3.7 \pm 1.1$, calculated at $z = 0.171$) and the Doppler factor, we estimate the bulk Lorentz factor, $\Gamma=8.3$, and the inclination angle of the jet with respect to the line of sight, $\phi=28.5 \degr$ \citep[see e.g.][]{1995PASP..107..803U}. Therefore the jet beaming parameters assuming $z_2=0.171$ are consistent with what is usually known for radio quasars, unlike the case if the source is at $z_1 = 5.063$.
The spectral index is $\alpha^{8\,\rm{GHz}}_{2\,\rm{GHz}} = -0.17 \pm 0.02 $ ($\alpha$ is defined as $S_\nu \propto \nu^\alpha$), indicating a flat radio spectrum at GHz frequencies. The beaming properties and radio spectral index of J2346+0705 classify it as a typical flat-spectrum radio-loud quasar.

\subsection{Applicability of the method to other objects}

To apply our method to other objects, it is required that a candidate high-redshift AGN is radio-loud and its prominent mas-scale radio jet structure is imaged with VLBI at multiple epochs (at least twice) at the same frequency. While these requirements obviously limit the widespread use of checking the high-redshift identification using jet proper motion data, there are in fact other suitable candidates found in the SDSS catalogues with VLBI data available. For example, the source J1110+4817 (SDSS J111036.32+481752.3)
is listed with $z = 6.168$ in SDSS DR13\footnote{\url{http://skyserver.sdss.org/dr13/en/tools/explore/Summary.aspx?id=1237658612517307020}}. However, \citet{1996MNRAS.282.1274H} and \citet{2016MNRAS.462.1603Y} independently gave $z = 0.74$. The currently available VLBI imaging data allow us to model the jet components at 8.7~GHz and estimate their apparent proper motions \citep{2020MNRAS.496.1811K}. However, this source does not show significant proper motion \citep{2020MNRAS.496.1811K}. Moreover, its radio structure resembles that of the compact symmetric objects (CSOs) which often show slow jet motions \citep{2012ApJ...760...77A,2012ApJS..198....5A}.
Another quasar, J2253+1942 (SDSS J225307.36+194234.6) has a very high redshift of $z_2 = 5.936$ in SDSS DR14\footnote{\url{http://skyserver.sdss.org/dr14/en/tools/explore/Summary.aspx?id=1237679504848912411}}, however, earlier literature data \citep{1998A&AS..128..507E} as well as SDSS DR16\footnote{\url{http://skyserver.sdss.org/dr16/en/tools/explore/Summary.aspx?id=1237679504848912411}} indicate a much lower value, $z = 0.284$. Plenty of archival VLBI imaging observations are available for this object. However, its compact, nearly featureless mas-scale radio structure is not well suited for identifying jet components and measuring their proper motions.
The studies of J1110+4817 and J2253+1942 illustrate the limitations of our proposed method for sources with unbeamed relativistic jets, or beamed sources without prominent jet components. 

Our proposed jet proper motion--based method to check the high-redshift identification of certain AGNs could be applied for more cases in the future when photometric redshift determinations for massive survey data are expected to reach out to much higher redshifts than today \citep[e.g.][]{2020Ap&SS.365...50R}.

\section{Summary}
\label{summary}

We described a method based on multiple-epoch VLBI imaging of jetted radio AGNs to check the validity of extremely high spectroscopic redshift measurements. The method rests on VLBI studies of large samples that indicate a well-defined upper bound of the apparent proper motion--redshift relation for pc-scale AGN jet components. This is a combined effect of jet physics with a maximum bulk Lorentz factor of the plasma, and the cosmological time dilation in the expanding Universe. If an apparent jet component speed exceeding about $40\,c$ is found in a source at any redshift, the object is very likely located at lower redshift. As an example, we analysed the jet properties of a suspicious high-redshift radio-loud AGN, J2346+0705. It has ambiguous redshift values reported in the literature. The inferred fast jet component proper motion in J2346+0705 excludes that it is a high-redshift ($z > 5$) object. Its radio spectrum and relativistic beaming parameters make it consistent with a flat-spectrum radio-loud quasar at $z = 0.171$.

Albeit with limitations, the proposed method could be applied to check other AGNs with ambiguous very high redshift identifications. As spectroscopic surveys advance and reach fainter magnitudes in the future, the automatic emission line identification and redshift determination algorithms will undoubtedly lead to more and more cases for uncertain or ambiguous redshift determination. Although limited in its scope because of the need for VLBI-monitored jetted radio AGN with fast component motions, the method presented here can dismiss certain cases of false high redshift measurements.

\section*{Data availability}

The datasets underlying this article were derived from sources in the public domain as given in the respective footnotes.

\section*{Acknowledgements}

This work is supported by the National Key R\&D Programme of China (2018YFA0404603), and the Hungarian National Research, Development and Innovation Office (grant 2018-2.1.14-T\'ET-CN-2018-00001). The authors acknowledge the use of archival calibrated VLBI data from the Astrogeo Center database maintained by Leonid Petrov. The National Radio Astronomy Observatory is a facility of the National Science Foundation operated under cooperative agreement by Associated Universities, Inc. This research has made use of the NASA/IPAC Extragalactic Database (NED) which is operated by the Jet Propulsion Laboratory, California Institute of Technology, under contract with the National Aeronautics and Space Administration. 





\begin{thebibliography}{99}

\bibitem[\protect\citeauthoryear{Abdo, et al.}{2010}]{2010ApJS..188..405A} Abdo A.~A., et al., 2010, ApJS, 188, 405
\bibitem[\protect\citeauthoryear{Acero et al.}{2015}]{2015ApJS..218...23A} Acero F., et al., 2015, ApJS, 218, 23
\bibitem[\protect\citeauthoryear{Ackermann et al.}{2013}]{2013ApJS..209...34A} Ackermann M., et al., 2013, ApJS, 209, 34
\bibitem[\protect\citeauthoryear{Ackermann et al.}{2017}]{2017ApJ...837L...5A} Ackermann M., et al., 2017, ApJL, 837, L5
\bibitem[\protect\citeauthoryear{Ahumada, et al.}{2020}]{2020ApJS..249....3A} Ahumada R., et al., 2020, ApJS, 249, 3
\bibitem[\protect\citeauthoryear{Ajello et al.}{2017}]{2017ApJS..232...18A} Ajello M., et al., 2017, ApJS, 232, 18
\bibitem[\protect\citeauthoryear{Albareti et al.}{2017}]{2017ApJS..233...25A} Albareti F.~D., et al., 2017, ApJS, 233, 25
\bibitem[\protect\citeauthoryear{Alexandroff et al.}{2013}]{2013MNRAS.435.3306A} Alexandroff R., et al., 2013, MNRAS, 435, 3306
\bibitem[\protect\citeauthoryear{An \& Baan}{2012}]{2012ApJ...760...77A} An T., Baan W.~A., 2012, ApJ, 760, 77
\bibitem[\protect\citeauthoryear{An et al.}{2012}]{2012ApJS..198....5A} An T., et al., 2012, ApJS, 198, 5
\bibitem[\protect\citeauthoryear{An et al.}{2020}]{2020NatCo..11..143A} An T., et al., 2020, Nature Communications, 11, 143
\bibitem[\protect\citeauthoryear{Ba{\~n}ados et al.}{2016}]{2016ApJS..227...11B} Ba{\~n}ados E., et al., 2016, ApJS, 227, 11
\bibitem[\protect\citeauthoryear{Beasley et al.}{2002}]{2002ApJS..141...13B} Beasley A.~J., Gordon D., Peck A.~B., Petrov L., MacMillan D.~S., Fomalont E.~B., Ma C., 2002, ApJS, 141, 13
\bibitem[\protect\citeauthoryear{Becker, White \& Edwards}{1991}]{1991ApJS...75....1B} Becker R.~H., White R.~L., Edwards A.~L., 1991, ApJS, 75, 1
\bibitem[\protect\citeauthoryear{Bolton et al.}{2012}]{2012AJ....144..144B} Bolton A.~S., et al., 2012, AJ, 144, 144
\bibitem[\protect\citeauthoryear{Britzen et al.}{2008}]{2008A&A...484..119B} Britzen S., et al., 2008, A\&A, 484, 119
\bibitem[\protect\citeauthoryear{Cohen et al.}{1988}]{1988ApJ...329....1C} Cohen M.~H., Barthel P.~D., Pearson T.~J., Zensus J.~A., 1988, ApJ, 329, 1
\bibitem[\protect\citeauthoryear{Condon et al.}{1998}]{1998AJ....115.1693C} Condon J.~J., Cotton W.~D., Greisen E.~W., Yin Q.~F., Perley R.~A., Taylor G.~B., Broderick J.~J., 1998, AJ, 115, 1693
\bibitem[\protect\citeauthoryear{Engels et al.}{1998}]{1998A&AS..128..507E} Engels D., Hagen H.-J., Cordis L., K\"{o}hler S., Wisotzki L., Reimers D., 1998, A\&AS, 128, 507
\bibitem[\protect\citeauthoryear{Fan et al.}{2001}]{2001AJ....122.2833F} Fan X., et al., 2001, AJ, 122, 2833
\bibitem[\protect\citeauthoryear{Frey et al.}{2015}]{2015MNRAS.446.2921F} Frey S., Paragi Z., Fogasy J.~O., Gurvits L.~I., 2015, MNRAS, 446, 2921
\bibitem[\protect\citeauthoryear{Healey et al.}{2007}]{2007ApJS..171...61H} Healey S.~E., et al., 2007, ApJS, 171, 61
\bibitem[\protect\citeauthoryear{Homan et al.}{2015}]{2015ApJ...798..134H} Homan D.~C., et al., 2015, ApJ, 798, 134
\bibitem[\protect\citeauthoryear{Hook et al.}{1996}]{1996MNRAS.282.1274H} Hook I.~M., McMahon R.~G., Irwin M.~J., Hazard C., 1996, MNRAS, 282, 1274
\bibitem[\protect\citeauthoryear{Jiang et al.}{2016}]{2016ApJ...833..222J} Jiang L., et al., 2016, ApJ, 833, 222
\bibitem[\protect\citeauthoryear{Kellermann et al.}{2004}]{2004ApJ...609..539K} Kellermann K.~I., et al., 2004, ApJ, 609, 539
\bibitem[\protect\citeauthoryear{Krezinger, et al.}{2020}]{2020MNRAS.496.1811K} Krezinger M., Frey S., An T., Jaiswal S., Zhang Y., 2020, MNRAS, 496, 1811
\bibitem[\protect\citeauthoryear{Lister et al.}{2016}]{2016AJ....152...12L} Lister M.~L., et al., 2016, AJ, 152, 12
\bibitem[\protect\citeauthoryear{Lister et al.}{2019}]{2019ApJ...874...43L} Lister M.~L., et al., 2019, ApJ, 874, 43
\bibitem[\protect\citeauthoryear{Paiano et al.}{2017}]{2017ApJ...851..135P} Paiano S., Falomo R., Franceschini A., Treves A., Scarpa R., 2017, ApJ, 851, 135
\bibitem[\protect\citeauthoryear{Perger et al.}{2018}]{2018MNRAS.477.1065P} Perger K., et al., 2018, MNRAS, 477, 1065
\bibitem[\protect\citeauthoryear{Perger et al.}{2019}]{2019MNRAS.490.2542P} Perger K., Frey S., Gab{\'a}nyi K. {\'E}., T{\'o}th L.~V., 2019, MNRAS, 490, 2542
\bibitem[\protect\citeauthoryear{Pushkarev \& Kovalev}{2015}]{2015MNRAS.452.4274P} Pushkarev A.~B., Kovalev Y.~Y., 2015, MNRAS, 452, 4274
\bibitem[\protect\citeauthoryear{Shen et al.}{2019}]{2019ApJ...873...35S} Shen Y., et al., 2019, ApJ, 873, 35
\bibitem[\protect\citeauthoryear{Shepherd}{1997}]{1997ASPC..125...77S} Shepherd M.~C., 1997, in Hunt G., Payne H.~E., eds, Astronomical Data Analysis Software and Systems VI, ASP Conf. Ser. 125. Astron. Soc. Pac., San Francisco, p. 77
\bibitem[\protect\citeauthoryear{Urry \& Padovani}{1995}]{1995PASP..107..803U} Urry C.~M., Padovani P., 1995, PASP, 107, 803
\bibitem[\protect\citeauthoryear{Readhead}{1994}]{1994ApJ...426...51R} Readhead A.~C.~S., 1994, ApJ, 426, 51
\bibitem[\protect\citeauthoryear{Reza \& Haque}{2020}]{2020Ap&SS.365...50R} Reza M., Haque M.~A., 2020, Ap\&SS, 356, 50
\bibitem[\protect\citeauthoryear{Vermeulen \& Cohen}{1994}]{1994ApJ...430..467V} Vermeulen R.~C., Cohen M.~H., 1994, ApJ, 430, 467
\bibitem[\protect\citeauthoryear{Volonteri et al.}{2011}]{2011MNRAS.416..216V} Volonteri M., Haardt F., Ghisellini G., Della Ceca R., 2011, MNRAS, 416, 216
\bibitem[\protect\citeauthoryear{Yuan, Strauss \& Zakamska}{2016}]{2016MNRAS.462.1603Y} Yuan S., Strauss M.~A., Zakamska N.~L., 2016, MNRAS, 462, 1603
\bibitem[\protect\citeauthoryear{Zhang, An \& Frey}{2020}]{2019arXiv191212597Z} Zhang Y., An T., Frey S., 2020, Science Bulletin, 65, 525


\end{thebibliography}





\bsp	
\label{lastpage}
\end{document}